\documentclass[12pt]{article}
\usepackage{apacite}
\usepackage{graphicx}
\usepackage{rotating}
\usepackage{dcolumn}
\usepackage{amssymb}
\usepackage{epsfig}
\usepackage{amsmath,amssymb,amsfonts,textcomp}
\usepackage{color}

\begin{document}

\setcounter{page}{1}
\setcounter{footnote}{0}
\topmargin 0pt
\oddsidemargin 5mm
\begin{titlepage}

\vspace{0.5cm}
\begin{center}
{\large {\bf Recording from two neurons: second order stimulus reconstruction from spike trains and population coding}}\\
\vspace{1.8cm}
{\large
{\bf N. M. Fernandes, B. D. L. Pinto, L. O. B. Almeida, J. F. W. Slaets}
and
 {\bf R. K\"oberle}\\
\indexspace
{\em Depart. de F\'{\i}sica e Inform\'atica, Inst. de F\'{\i}sica de S\~ao Carlos,
USP, C. P. 369, CEP 13560-970, S\~ao Carlos-SP, Brasil, e-mail: rk@if.sc.usp.br}}\\

\vspace{1cm}
\today
\end{center}

{\bf
We study the reconstruction of  visual stimuli from spike trains, recording simultaneously from
the two H1 neurons located in the lobula plate of the fly {\em Chrysomya megacephala}.
The fly views two types of stimuli, corresponding to rotational and translational displacements.
If the reconstructed stimulus is to be represented by  a Volterra series
and  correlations between spikes are to be taken into account, first order expansions are insufficient
and we have to go to second order, at least.
In this case higher order correlation functions have to be manipulated, whose size may become
prohibitively large. We therefore develop a Gaussian-like representation for fourth order correlation
functions, which works exceedingly well in the case of the fly. The reconstructions using this Gaussian-like representation are very similar to  the  reconstructions using the experimental correlation functions.
The overall contribution to rotational stimulus reconstruction of  the second order kernels - measured by a chi-squared averaged over the whole experiment -   is
only about 8\% of the first order contribution.
Yet if we introduce an instant-dependent chi-square to measure the contribution of second order kernels
at special events, we observe an up to  100\% improvement.
As may be expected, for translational stimuli the reconstructions are rather poor.
The Gaussian-like representation could be a valuable aid in
 population coding with large number of neurons.
}
\end{titlepage}

\newpage
\section{Introduction \hrulefill}
Living animals have to reconstruct a representation of the external world from the output of their sensory systems in order to correctly react to
the demands of a rapidly varying
environment. In many cases this sensory output is encoded into a sequence of identical action potentials, called spikes.
If we represent the external world by a time-dependent stimulus function $s(t)$, the  animal has to reconstruct $s(t)$ from a set of spikes.
This decoding procedure  generates
an estimate $s_{e}(t)$ of the stimulus like a digital-to-analog converter.

\begin{figure}[!ht]
  \centerline{\hbox{\psfig{file=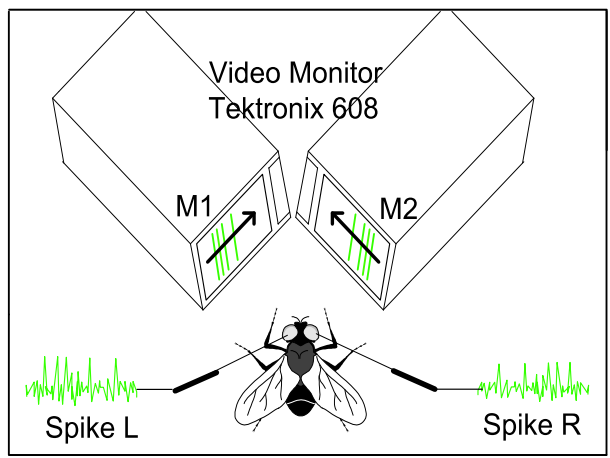,height=6cm,width=8cm}}}
  \caption{ {\sf  Motion sensitivity of the two H1 neurons. Each eye sees a monitor displaying a rigidly moving bar pattern. The stimuli in this figure correspond to a translational motion in which both neurons are excited. Inverting the stimulus shown  by monitor M1 would generate a rotational stimulus, which now inhibits the response of the left neuron. Electrodes record extracellularly from each H1. }}
\label{setup}
\end{figure}

\begin{figure}[!ht]
\centerline{\hbox{\psfig{file=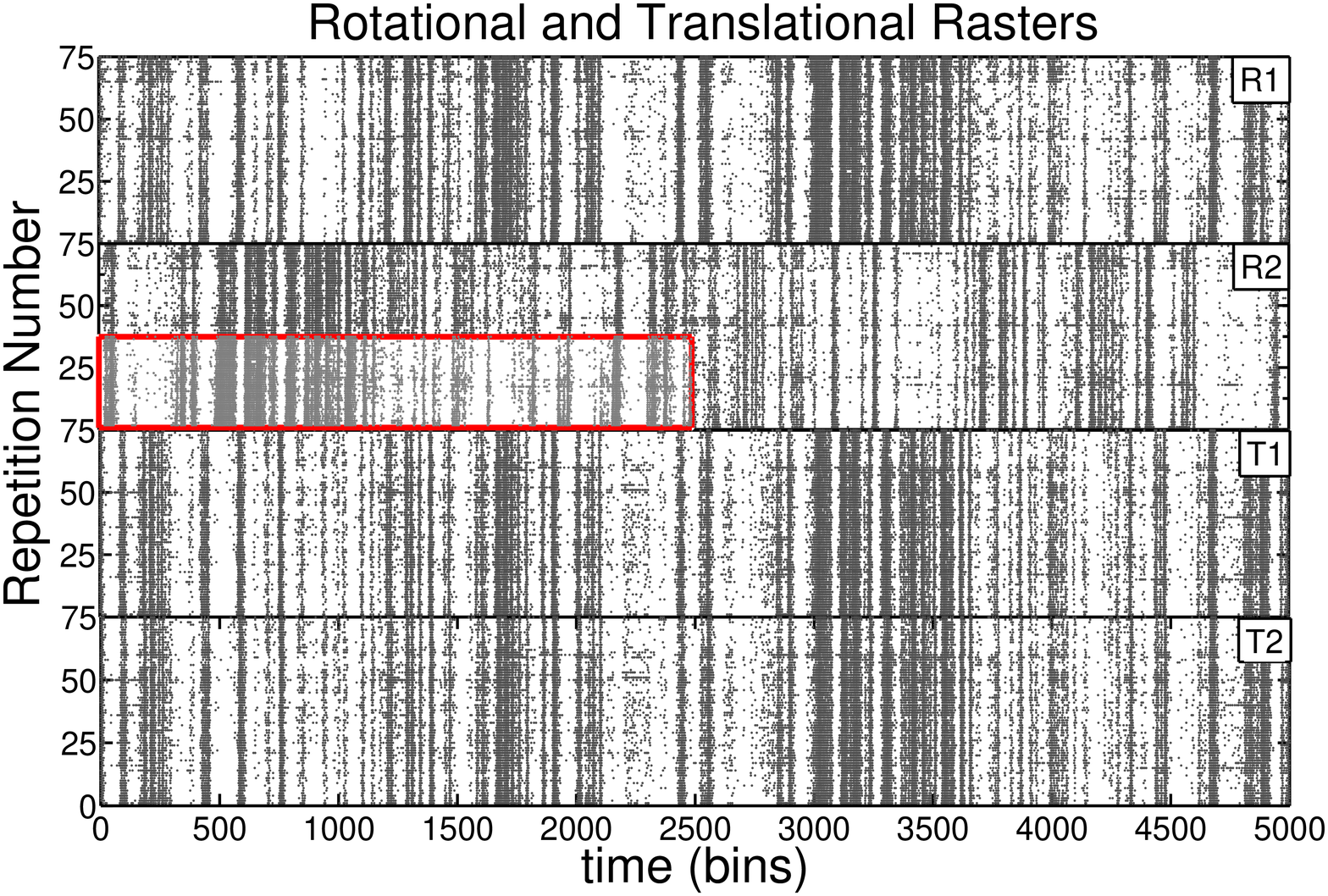,height=12cm,width=13cm}}}  \caption{ {\sf  Raster plot for the two H1 neurons, showing their complementary action under rotational and translational stimuli. The same time-dependent stimulus $s(t)$ is  repeatedly shown to the fly, the horizontal  time axis running from time zero to 5000 bins = 10 seconds and the vertical axis showing the repetition number. The responses of the neurons are shown as a raster, where each dot represents a  spike.
The right H1 sees a stimulus $s_r(t)$ and the left one sees  $s_l(t)$.
Rotational stimuli $s_r(t)=s_l(t)=s(t)$: (R1) spikes from right H1 and (R2) spikes from left H1.
Translational stimuli $s_r(t)=-s_l(t)=s(t)$: (T1) spikes from right H1 and (T2) spikes from left H1.
Inset to (R2): Spikes from right H1, fly subjected to sign reversed stimuli in order to simulate raster (R2).
}}
\label{raster}
\end{figure}

Here we study this decoding procedure in a prominent example of spiking neurons:
the two H1 neurons of the fly {\it Chrysomya megacephala}.
The fly has two compound eyes with their associated neural processing systems \cite{hausen:1981,hausen:1982b,hausen:1984}.
Motion detection starts at the photoreceptor cells, eight of them located in each one of the $\sim 5000$ ommatidia of each compound eye. They effect the transduction of photons into electrical signals,
which are propagated via the lamina and medulla to the lobula plate. This neuropil is - inter alia - composed of horizontally and vertically directionally  sensitive  wide field neurons. The H1 neurons are
horizontally sensitive and are
excited by ipsilateral back to front motion and inhibited by  oppositely moving stimuli. Each H1 neuron projects its axon to the contralateral lobula plate, exciting there two horizontal and two
centrifugal cells. These cells mediate mutual inhibition between the two H1 neurons \cite{Haag1999,Haag2001,Farrow2003,Haag2008,Krapp2009}\footnote{Although experimental work has focussed on the vertical system, one expects analog results for the horizontal one.}.
We subject the fly to rotational and translational stimuli - see Figure \ref{setup}.
If the fly rotates around a vertical axis, say clockwise when looking down the axis, the left  neuron
is inhibited and the right one is exited,
 so that the two neurons become an efficient rotational detector \cite{hausen:1984}.
This can be seen in Figure \ref{raster} (R1) \& (R2). Even when recording
 only from the ipsilateral  H1, one can simulate the response of the contralateral H1. In fact, since the two H1 cells have mirror symmetric
directional sensitivities, the sign flipped stimulus induces a response in the ipsilateral H1 typical for the contralateral H1 cell \cite{Spikes}. The inset in (R2) shows this to be true to a very good approximation.

In forward translation none of  H1 neurons is excited, corresponding to the low spike density regions in the raster-plots of
Figure \ref{raster} (T1) \& (T2).
In backward translation, both H1's are excited and we expect a strong inhibition. Yet the spike rate  is comparable to rotational excitation - compare Figure \ref{raster} (R1) \& (T1).  Numerical computation  confirms this  visual impression. Nevertheless
in translation the two H1's fire mainly in sync, which leads to subtle differences with respect to rotation. As a consequence, our reconstructions will be much poorer for the translational case - see section \ref{sect:reconstr}.

If we want to take correlations between spikes into account, instead of treating them independently, we have to go at least to second order stimulus reconstructions.
These  require the computation of higher order spike-spike correlation functions and a subsequent matrix inversion.  If one records from many neurons simultaneously, the size of these matrices may soon become prohibitively large.
Here we present an efficient representation of these higher order correlation functions in terms of second order ones. The reconstruction now costs far less computationally, avoids large matrix inversions and gives excellent results. We test the quality of our reconstructions under both rotational and  translational stimuli.

If this representation holds more generally, it may well make population coding computationally
 more tractable. We briefly discuss a perturbation scheme, which allows a stepwise inclusion of small effects.

\section{Stimulus reconstruction from spike trains \hrulefill}
\label{reconstr}

Suppose we want to reconstruct the stimulus from the response of a single H1 neuron. We represent this response as a  spike train  $\rho(t)=\sum_{i=1}^{N_s} \delta(t-t_i)$,
which is a sum of delta functions at the spike times $t_i$. $N_s$ is the
total number of spikes generated by the neuron during the experiment.

The simplest reconstruction extracts the stimulus estimate via a linear transformation, see e.g.  \cite{Spikes,Bialek:1991},

\begin{equation}
\label{vol1}
	s_e(t) = \int_{-\infty}^{\infty} k_1(\tau)\rho(t-\tau)d\tau,
\end{equation}
with the kernel $k_1(t)$ to be determined.

For simplicity we effect an {\em acausal} reconstruction, i.e. we integrate from
$-\infty$ to $+\infty$. Essentially the same results are obtained in a causal
reconstruction. One way to implement causality proceeds to  estimate the stimulus at time $t$, using as input the spike train up to time $t+t_0$. For the fly $t_0$  has to be $\gtrsim$  to 30 milliseconds.
In this case equation \ref{vol1} would read:
$	s_e(t) = \int^{\infty}_{-t_0} k_1(\tau)\rho(t-\tau)d\tau.$

Equation \ref{vol1} is the first term of a Volterra series   \cite{volterra}:
\begin{equation}
\label{vol2}
		s_e(t) = \int_{-\infty}^{\infty} k_1(\tau)\rho(t-\tau)d\tau+
	\int_{-\infty}^{\infty} k_2(\tau_1,\tau_2)\rho(t-\tau_1)\rho(t-\tau_2)d\tau_1 d\tau_2 +\ldots
\end{equation}
There is no convergence proof for this expansion, but heuristically we
may say that it should be a valid approximation, if the average number of spikes per correlation time $\tau_c$,
\begin{equation}
	\eta=\langle r \rangle \tau_c,
\end{equation}
is small \cite{Spikes}. Here $ \langle r \rangle$ is the mean spike rate and $\tau_c$ a typical signal correlation time. For small $\eta$ each spike gives independent information about the stimulus. In our case $\eta\sim 0.6-0.8$, which is of the order of  unity, so that higher order effects might be relevant.

The first order term, being  proportional to $\sum_{i}^{N_s}k_1(t-i_i)$, independently adds contributions for each spike. Yet it is well established that pairs of spikes carry a significant amount
of additional  information beyond the single spike contributions \cite{brenner1}. This motivates the
addition of the second order kernel $k_2(\tau_1,\tau_2)$, which includes
correlations between up to two spikes.

In order to obtain the  kernels $k_1$ and $k_2$ we choose to minimize the following functional - the $\chi^{(2)}$ error -
\begin{equation}
\label{chi}
	\chi^{(2)}(k_1,k_2) = \langle\int dt [s_e(t)-s(t)]^2 \rangle.
\end{equation}
	The brackets stand for an ensemble average with respect to the
distribution of all possible stimuli  in a given experiment.
In a long experiment we average over $N_w\sim 10^5$ time windows of size $T_w$. Typically $T_w\sim 100$ milliseconds - see section \ref{MM} for details.
For ease of presentation, in the following our discussions will always refer to the rotational setup, unless explicitly stated
otherwise as in section \ref{sect:reconstr}.

Since the functional \ref{chi} is quadratic, the equations minimizing $\chi^{(2)}(k_1,k_2)$

\begin{equation}
\label{dchi}
	\partial \chi^{(2)}/\partial k_j = 0, j=1,2
\end{equation}
are linear in the unknowns $k_1,k_2$. E.g., if we keep only $k_1$, using therefore equation \ref{vol1}, we get:

\begin{equation}
\label{k11}
	\tilde k_1(\omega) = \frac{ \langle\tilde s(\omega)^* \tilde\rho(\omega) \rangle}{ \langle\tilde\rho(\omega)^*\tilde\rho(\omega) \rangle},
\end{equation}
where Fourier transforms are defined as
$\tilde F(\omega) = \int dt  F(t) e^{\imath\omega t}$.

We may include the second order term $k_2$, either as a correction to the first order reconstruction $s_1(t) = k_1\star\rho(t)$
\footnote{The symbol $\star$ stands for a convolution as in equation \ref{vol1}.},
 or one may solve the coupled system \ref{dchi}.

If we record simultaneously from left and right H1, we obtain two spike trains
$\rho_1(t)$ and $\rho_2(t)$. The expansion equation \ref{vol2} generalizes to
\begin{equation}
\nonumber
\label{vol22}
s_e(t) = K_1\star\rho_1(t)+ K_2\star\rho_2(t)+
\end{equation}
\begin{equation}
K_{11}\star\rho_1\star\rho_1(t)+K_{12}\star\rho_1\star\rho_2(t)+K_{22}\star\rho_2\star\rho_2(t)+\ldots.
\end{equation}
Here we have included the kernel $K_{12}$, which  encodes effects correlating $\rho_1$ and $\rho_2$
\footnote
{Notice that we have not {\em orthogonalized} our expansion equation \ref{vol2}, so that
there are  $K_{11}(t_1,t_1)$ terms, which could have been absorbed in $K_1(t)$ and similarly for $K_2(t)$.}.
Notice that $K_{12}=K_{21}$.

To first order, keeping only $K_{1}$ and $K_{2}$ in the expansion \ref{vol22},  we get the following equations:
\begin{equation}
       \widetilde{S\!\!R}_a(\omega) = \sum_{b=1}^2 \tilde K_b(\omega) \tilde R_{ab}(\omega),\,\, a=1,2
\label{K1}
\end{equation}
where
\begin{equation}
	\widetilde{S\!\!R}_a(\omega) = \int dt dt' \langle s(t')\rho_a(t'-t) \rangle e^{\imath\omega t}.
\end{equation}
 and
\begin{equation}
       R_{ab}(t_1,t_2) =\int dt  \langle\rho_a(t-t_1)\rho_b(t-t_2) \rangle,\; a,b=1,2.
\end{equation}
Due to time-translation invariance $R_{ab}(t_1,t_2)$ is only a function of the difference: $R_{ab}(t_1,t_2)=R_{ab}(t_1-t_2)$ and $ \tilde R_{ab}(\omega) = \int dt  R_{ab}(t) e^{\imath\omega t}$. Analogous properties hold for  all the
following correlation functions involving only $\rho(t)$.

The  solution of equations \ref{K1} yields
\begin{equation}
      {\tilde K}_a(\omega) = (L_a(\omega) R_{\hat{a}\hat{a}}-L_{\hat{a}}(\omega)R_{a \hat{a}}(\omega) )/\Delta,\; a=1,2
\end{equation}
where
\begin{equation}
	L_a(\omega) =  \langle s(\omega)\rho^*_a(\omega) \rangle,\\
	\Delta = R_{11}R_{22}-R_{12}R_{21}
\end{equation}
and $\hat{a}=3-a$.
We obtain the first order reconstruction as
\begin{equation}
       s_1(t) = K_1\star\rho_1(t)+K_2\star\rho_2(t).
\end{equation}

 Since the second order contribution turns out to be small, we treat it as a perturbation to the first order reconstruction.
We therefore expand   $s_2(t) = s(t)-s_1(t)$ as:
\begin{equation}
s_2(t) = K_{11}\star\rho_1\star\rho_1(t)+K_{12}\star\rho_1\star\rho_2(t)+K_{22}\star\rho_2\star\rho_2(t).
\end{equation}

We now have to solve  the following equations
\begin{equation}
S\!\!R^{(2)}_{ab}(t_1,t_2)=
	 \int dt_3 dt_4  \sum_{c,d=1}^2 K_{cd}(t_1,t_2) R^{(4)}_{abcd}(t_1,t_2,t_3,t_4),
\label{K22}
\end{equation}
where
\begin{equation}
 S\!\!R^{(2)}_{ab}(t_1,t_2) = \int dt  \langle s_2(t)\rho_a(t-t_1)\rho_b(t-t_2) \rangle,
\end{equation}
\begin{equation}
 R^{(4)}_{abcd}(t_1,t_2,t_3,t_4) =\int  dt  \langle\rho_a(t-t_1)\rho_b(t-t_2)\rho_c(t-t_3)\rho_d(t-t_4) \rangle.
\end{equation}

Although the system  \ref{K22} is linear, the matrices to be inverted may be very large. We have to invert the matrix
\begin{equation}
 {\cal M}_{AT}^{BT'}\equiv R^{(4)}_{abcd}(t_1,t_2,t_3,t_4),
\label{4ptmat}
\end{equation}
where $A,B$ are compound indices $A=[ab],B=[cd]$ labeling the neurons.  $T=[t_1,t_2],T'=[t_3,t_4]$ are compound time indices of size $T_w^2$ each. If we compute the correlation functions using a time window of $T_w=128$  bins, with binsize $=2$ milliseconds, then the size of ${\cal M}_{AN}^{BN'} $ is  $\sim 128^4\times2^4 \sim  5\times 10^9$. The matrices to be inverted may become prohibitively large, especially if we record from more than just two neurons
\footnote{
We may solve the above system in Fourier space and select a subset of frequencies in order to reduce the size of the system.
}.

We therefore present below a Gaussian-like representation of  $R^{(4)}_{abcd}$  with a small number of parameters and which requires no large matrix inversion.

\section{Gaussian-like (Gl) representation for  4-point functions\hrulefill}
\label{1G}

\begin{figure}[!ht]
  \centerline{\hbox{\psfig{file=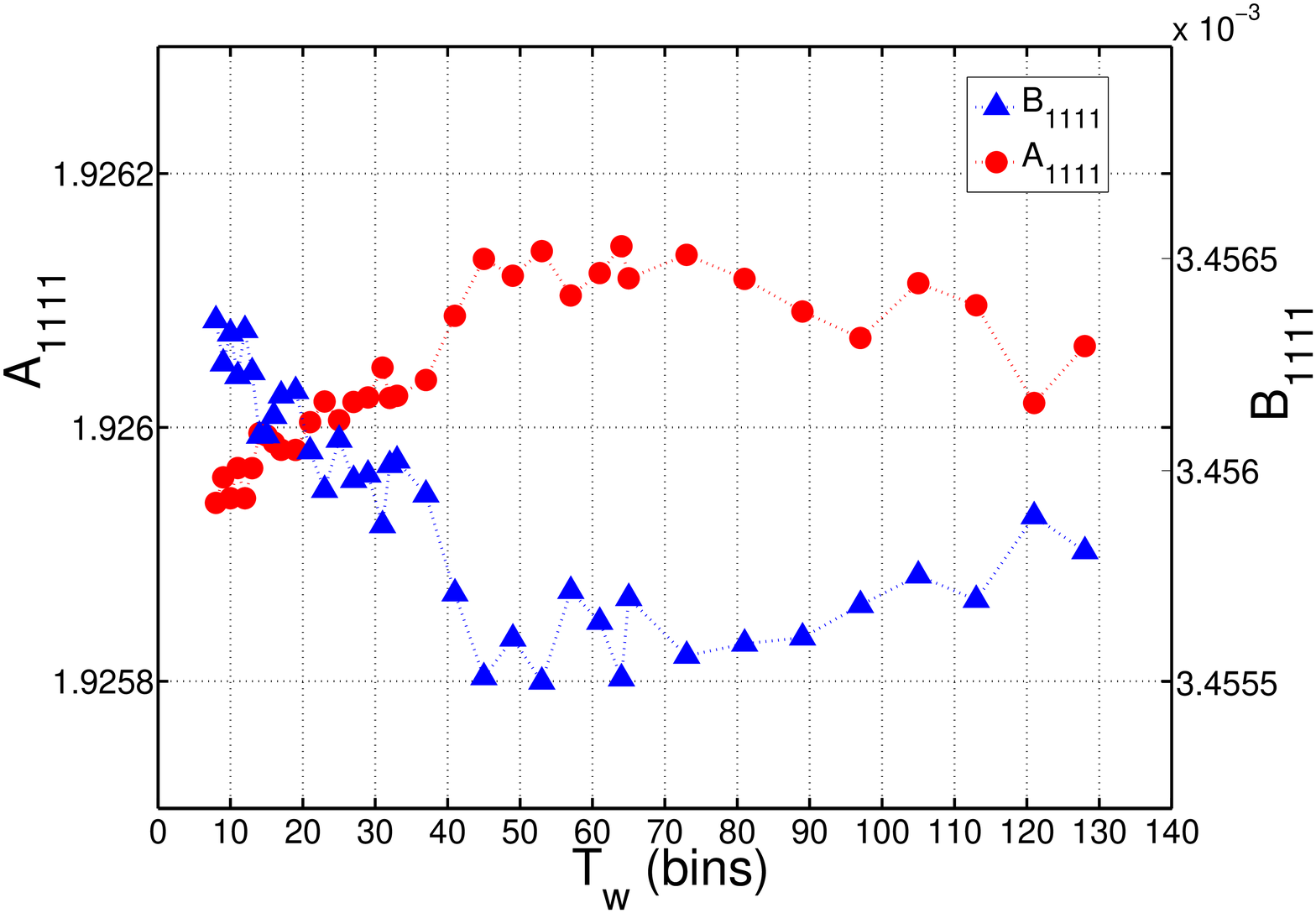,height=10cm,width=12cm}}}
  \caption{ {\sf  Window-size dependence of parameters  $A_{1111}$ and $B_{1111}$. Similar behavior is found for the other parameters $A$ and $B$. Notice that variations are on the $0.05$\% level.}}
\label{AB_dep}
\end{figure}

\begin{figure}[!ht]
  \centerline{\hbox{\psfig{file=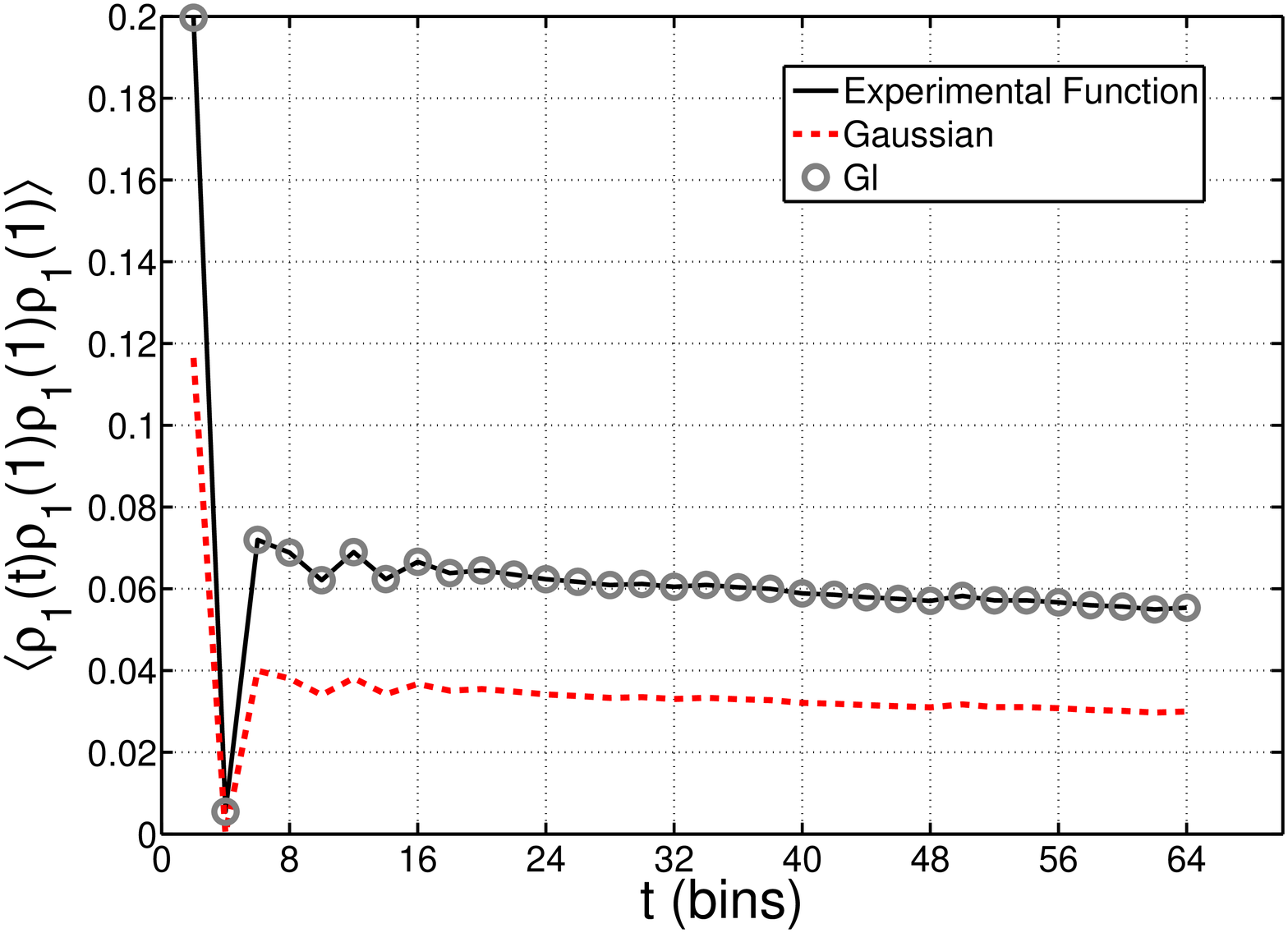,height=8cm,width=12cm}}}
  \caption{ {\sf  4-point functions and its  Gl approximation.
We plot  $R_{1111}(t_1,t_2=t_3=t_4=1)$ for window-size $T_w=64$ bins.
The black continuous line is the experimental 4-point function.
 The dashed line is its Gaussian approximation without parametrization using
equation \ref{GR4}.
The circles represent its Gl approximation \ref{Fit1}.
}}
\label{Fig:4_pt_fits1}
\end{figure}

\begin{figure}[!ht]
  \centerline{\hbox{\psfig{file=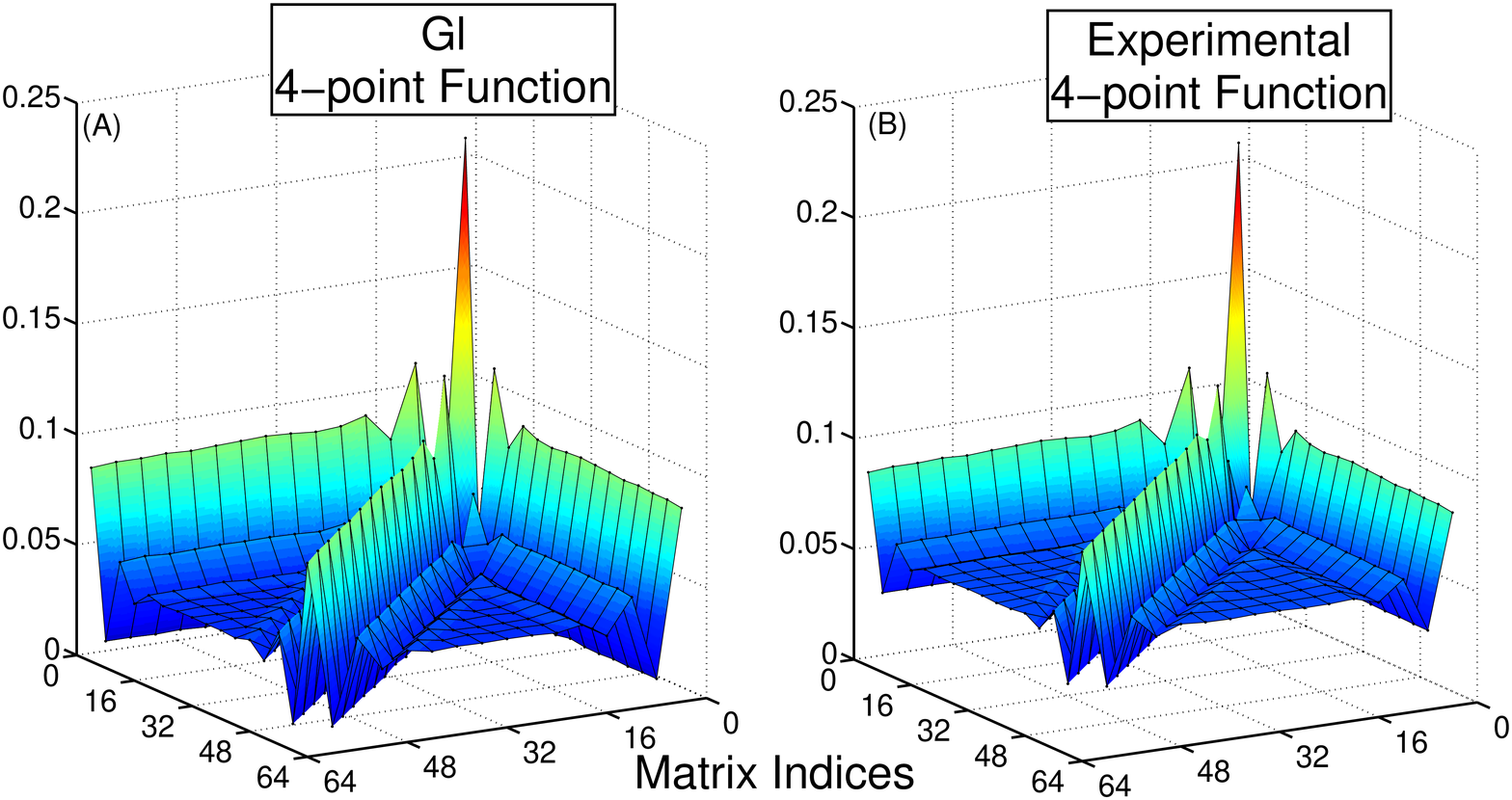,height=12cm,width=15cm}}}
 \caption{ {\sf  (A) Gl approximation  and (B) experimental  4-point function for  $R_{1111}(t_1,t_2,t_3=t_4=1)$ for window-size  $T_w=64 $ bins.
}}
\label{Fig:4pt_mat}
\end{figure}

In this section we present a representation of the 4-point function ${ R}^{(4)}_{abcd}$ in
terms of the 2-point function ${ R}^{(2)}_{ab}$, which is surprisingly good and
which avoids the computation of the large matrices  \ref{4ptmat}.

If our spike-generating process were Gaussian, we would have
the following structure for $R^{(4)}$:
\begin{equation}
\nonumber
	R^{(4)}(1,2,3,4) = R(1,2)R(3,4)+R(1,3)R(2,4)+R(1,4)R(2,3)
\label{GR4}
\end{equation}
\begin{equation}
-2 \langle\rho(t) \rangle^4,
\end{equation}
where $\langle\rho(t) \rangle$ is just a constant, due to time-translation invariance\footnote{We write $(1,2,\ldots)$ instead of $(t_1,t_2,\ldots)$.}.

This suggests the following representation for $R^{(4)}$:
\begin{equation}
\nonumber
	R^{(4)}(1,2,3,4) = A \left[ R(1,2)R(3,4)+R(1,3)R(2,4)+\right.
\end{equation}
\begin{equation}
	\left. R(1,4)R(2,3) \right]-B,
\label{Fit1}
\end{equation}
where $A$ and $B$ are constants to be adjusted\footnote{Any structure built only from $R(t_1,t_2)$ could be used for our method to work.}.

For two neurons we get the  representation:
\begin{equation}
  R_{abcd}(1,2,3,4) =  [R_{ab}(1,2) R_{cd}(3,4) +
R_{ac}(1,3)  R_{bd}(2,4)+
\nonumber
\end{equation}
\begin{equation} R_{ad}(1,4)  R_{bc}(2,3)] A_{abcd}+B_{abcd}
\label{2NRep}
\end{equation}
with $a,b,c,d=1,2$ and $A_{abcd}$, $B_{abcd} $ constants to be determined.

The usefulness  of our Gl-representation scheme depends on the quality of the 4-point functions obtained, which in turn hinges on
the knowledge of  the constants $A_{abcd}$ and $B_{abcd}$.
There would be no point, if this required the computation  of 4-point functions in large window sizes and
a fitting procedure using these windows  - exactly what we wanted to avoid.
We therefore fit the constants $A_{abcd}$ and $B_{abcd}$ for a sequence of window sizes $T_w$, ranging from $10$ to $128$ bins, using
$  R_{1111}(t_1,t_2=t_3=t_4=1)$ to fit to the experimental data.
As can be seen in Figure \ref{AB_dep}, at least in the fly's case, the dependence of the  parameters $A_{abcd},B_{abcd}$ on $T_w$ is only
$0.05$\%  and therefore completely negligible.
The constants $A_{abcd}$ and $B_{abcd}$ can therefore be computed very fast  in small windows.
In Figure \ref{Fig:4_pt_fits1} we plot the fits to the first row
$ R_{1111}(t_1,t_2=t_3=t_4=1)$ and its Gl approximation.  As advertised we obtain a perfect fit.

In Figure \ref{Fig:4pt_mat} we show the Gl approximation for the
 $R_{1111}(t_1,t_2,t_3=t_4=1$ and its experimental version, which emphasizes the quality of the approximation.
Using the same parameters for the other entries of
$ R_{1111}$ and for $R_{2222}$ results in a fitting error  about 20 \% larger.

 One of the utilities of this representation will become apparent, once we deal with the solution of equation \ref{K22} in the next section.

\section{A convenient set of functions to solve for second order kernels\hrulefill}

At this point it is convenient to introduce a complete set of basis functions
 $f_\mu(t),\mu=1,2,..,n_f$ to expand our variables in. We thus trade continuous time-arguments for discreet  Greek indices.
We expand our second order kernels as:
\begin{equation}
	K_{ab}(t_1,t_2) = \sum_{\mu.\nu} \, f_\mu(t_1)f_\nu(t_2)  {\cal D}^{ab}_{\mu\nu}.
\end{equation}
We also expand our correlation functions:
\begin{equation}
S\!\!R^{(2)}_{ab}(t_1,t_2) = \sum_{\mu.\nu} {\cal S}^{ab}_{\mu\nu}\, \, f_\mu(t_1)f_\nu(t_2)
\label{expSR}
\end{equation}
and
\begin{equation}
\nonumber
R^{(4)}_{abcd}(t_1,t_2,t_3,t_4)=
\end{equation}
\begin{equation}
   \sum_{\alpha\beta\mu.\nu}  {\cal R}^{abcd}_{\alpha\beta\mu\nu}\,
f_\alpha(t_1)
f_\beta(t_2)f_\mu(t_3) f_\nu(t_4).
\label{expRRRR}
\end{equation}
\begin{figure}[!ht]
  \centerline{\hbox{\psfig{file=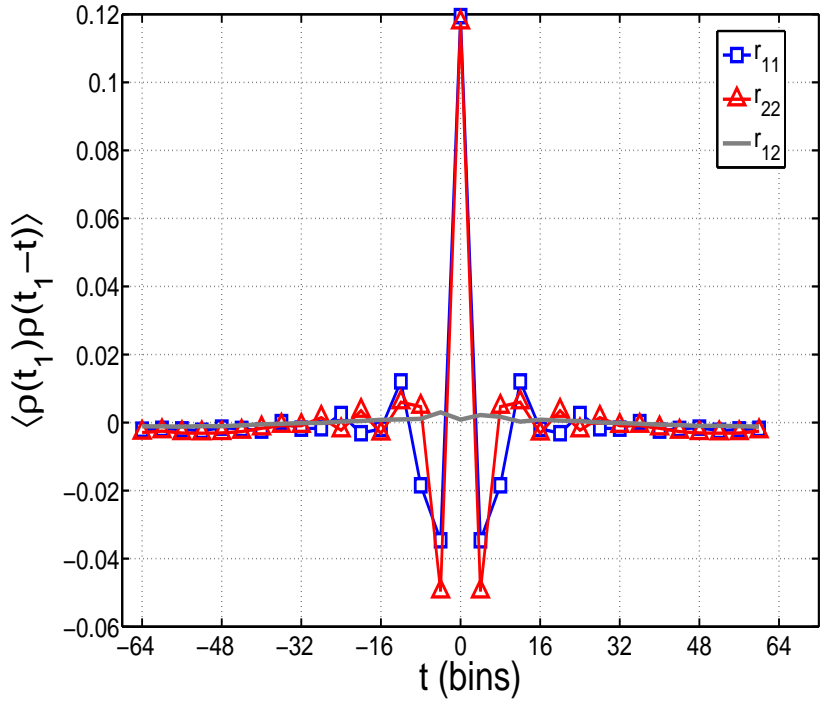,height=10cm,width=12cm}}}
  \caption{ {\sf 2-point correlation functions
$r_{11}(t)= \langle\rho_1(t_1-t)\rho_1(t_1) \rangle$, $ r_{22}(t)=\langle\rho_2(t_1-t)\rho_2(t_1) \rangle$ and $r_{12}(t)=\langle\rho_1(t_1-t)\rho_2(t_1) \rangle$. The central peak is absent in the mixed correlation function $r_{12}(t)$. }}
\label{R11_R22}
\end{figure}

In order to efficiently compute our second order kernels it is
crucial to select an adequate  set for
 $f_\mu(t),\mu=1,2,..,n_f$.

 Depending on the case, it may be sufficient to use a small number $n_f$ of functions $f_\mu(t)$ to get a useful representation. If $n_f$ has  only a slight dependence on
window size $T_w$, this would allow one to increase $T_w$ without further computational costs.

Often a Fourier expansion is used, i.e. $f_\omega=e^{\imath\omega t}$. But we may exploit our liberty to choose the functions in a more profitable way.
Since our 2-point function $R(t_1,t_2)$ is real, positive\footnote{In case this is not true, we just add a convenient constant.} and symmetric in $t_1,t_2$, it posses a complete set of eigenfunctions $h_\mu(t)$:
\begin{equation}
	\int dt_2 R(t_1,t_2) h_\mu(t_2) = r_\mu h_\mu(t_1)
\end{equation}
with eigenvalues $r_\mu, \mu=1,\ldots, N_w$.
We now choose our functions  as $f_\mu(t)=h_\mu(t)/\sqrt{r_\mu}$, which satisfy:
\begin{equation}
	\int dt_1 dt_2 f_\mu(t_1)R(t_1,t_2)f_\nu(t_2) = \delta_{\mu\nu}.
\end{equation}
This choice will avoid large matrix inversions, if at least part of our
higher order correlation functions can be built from $R(t_1,t_2)$.

Substituting the expansions \ref{expSR} and \ref{expRRRR}  into
 equations \ref{K22}, we get a  linear system to be solved for $ {\cal D}^{ab}_{\mu\nu}$:
\begin{equation}
\label{Lin}
{\cal S}_{ab}^{\mu\nu} =
\sum_{cd,\alpha\beta} {\cal R}_{abcd}^{\mu\nu\alpha\beta}\, {\cal D}_{cd}^{\alpha\beta}
\end{equation}

In order to avoid cluttering our expressions with indices, we introduce our representation first for one neuron only, suppressing thus the indices $a,b,..$, all set to $1$.
We choose our functions $f_\mu(t)$ to diagonalize
 $ R^{11}(t_1,t_2)={ \langle \rho_1(t_1)\rho_2(t_2)\rangle}$:
\begin{equation}
\int dt_1 dt_2 f_\mu(t_1)R^{11}(t_1,t_2)f_\nu(t_2) = \delta_{\mu\nu}.
\end{equation}
The first of equations \ref{2NRep} for  ${\cal R}_{\mu\nu\alpha\beta}^{1111}$ becomes
\begin{equation}
{\cal R}_{\mu\nu\alpha\beta} = A (\delta_{\mu\nu}\delta_{\alpha\beta}+
2\delta_{\mu\alpha}\delta_{\nu\beta})-2B\,n_\alpha n_\beta n_\mu n_\nu ,
\end{equation}
where $n_\mu = \int dt f_\mu(t) \langle\rho(t) \rangle$.

Using this expression
and the shorthand ${\cal S}_{\mu\nu}\equiv{\cal S}_{\mu\nu}^{11}$ in equations \ref{Lin}, we get the following
equations for the unknown coefficients ${\cal D}_{\mu\nu} \equiv{\cal D}_{\mu\nu}^{11}$
\begin{equation}
	{\cal S}_{\mu\nu} =A[ tr({\cal D})\delta_{\mu\nu}+2{\cal D}_{\mu\nu}]-2 B D_{nn} \,n_\nu n_\mu,
\label{Dequ1}
\end{equation}
where
$tr({\cal D})\equiv \sum_\mu{\cal D}_{\mu\mu} $
and
$ D_{nn}\equiv \sum_{\alpha\beta}n_\alpha {\cal D}_{\alpha\beta}n_\beta$.
The sums over $\mu,\alpha,\beta$ run from $1$ to $T_w$ bins.

This system can now easily be solved by:
\begin{enumerate}
\item
taking the trace over $\mu\nu$ to compute $tr({\cal D})\equiv D$  and
\item
multiplying by $n_\mu,n_\nu$ to compute $D_{nn}$.
\end{enumerate}
We get

\begin{equation}
	{\cal D}_{\mu\nu} = [{\cal S}_{\mu\nu}/A-D\,\delta_{\mu\nu}+2Bn_\mu n_\nu\, D_{nn}]/2,
\label{EquD1}
\end{equation}
with
\begin{equation}
	D = [2(1-n_4){\cal S}_{\mu\mu}+2n_2\,n_\mu{\cal S}_{\mu\nu} n_\nu]/\Delta,
\end{equation}
\begin{equation}
	D_{nn} = [(n+2)n_\mu {\cal S}_{\mu\nu} n_\nu - {\cal S}_{\mu\mu} n_2]/\Delta,
\end{equation}
where
\begin{equation}
	\Delta = 2(T_w+2)(1-n_4)+2n^2, n_2\equiv \sum_\mu n_\mu n_\mu, n_4 \equiv (n_2)^2.
\end{equation}

For two neurons we now have to decorate our formulas with the indices $a,b,\ldots$.
To simplify our formulas, we assume symmetry between the two neurons: $R_{11}=R_{22}$,
which in our case is very well satisfied - see Figure \ref{R11_R22}.

The 4-point functions are now represented as
\begin{equation}
\begin{array}{cc}
 {\cal R}_{1111}^{\mu\nu\alpha\beta} = & [\delta_{\mu\nu}\delta_{\alpha\beta} + \delta_{\mu\alpha}\delta_{\nu\beta}+\delta_{\mu\beta}\delta_{\nu\alpha}] A_{1111}+B_{1111}^{\alpha\beta\mu\nu}
\label{4ptrep2}
\\
 {\cal R}_{1112}^{\mu\nu\alpha\beta} =&  [\delta_{\mu\nu}R^{\alpha\beta}_{12}+\delta_{\mu\alpha}R^{\nu\beta}_{12}+\delta_{\nu\alpha}R^{\mu\beta}_{12}]A_{1112}+ B_{1112}^{\alpha\beta\mu\nu}
\\
 {\cal R}_{1122}^{\mu\nu\alpha\beta} =&  [\delta_{\mu\nu}\delta_{\alpha\beta}+R^{\mu\alpha}_{12}R^{\nu\beta}_{12}+R^{\mu\beta}_{12}R^{\nu\alpha}_{12}]A_{1122}+ B_{1122}^{\alpha\beta\mu\nu}
\\
 {\cal R}_{1222}^{\mu\nu\alpha\beta} =&  [R^{\mu\nu}_{12}\delta_{\alpha\beta}+R^{\mu\alpha}_{12}\delta_{\nu\beta}+R^{\mu\beta}_{12}\delta_{\nu\alpha}]A_{1222}+ B_{1222}^{\alpha\beta\mu\nu}
\\
{\cal R}_{2222}^{\mu\nu\alpha\beta}=& [\delta_{\mu\nu}\delta_{\alpha\beta} + \delta_{\mu\alpha}\delta_{\nu\beta}+\delta_{\mu\beta}\delta_{\nu\alpha}]
A_{2222}+B_{2222}^{\alpha\beta\mu\nu}.
\end{array}
\end{equation}

\begin{figure}[!ht]
 \centerline{\hbox{\psfig{file=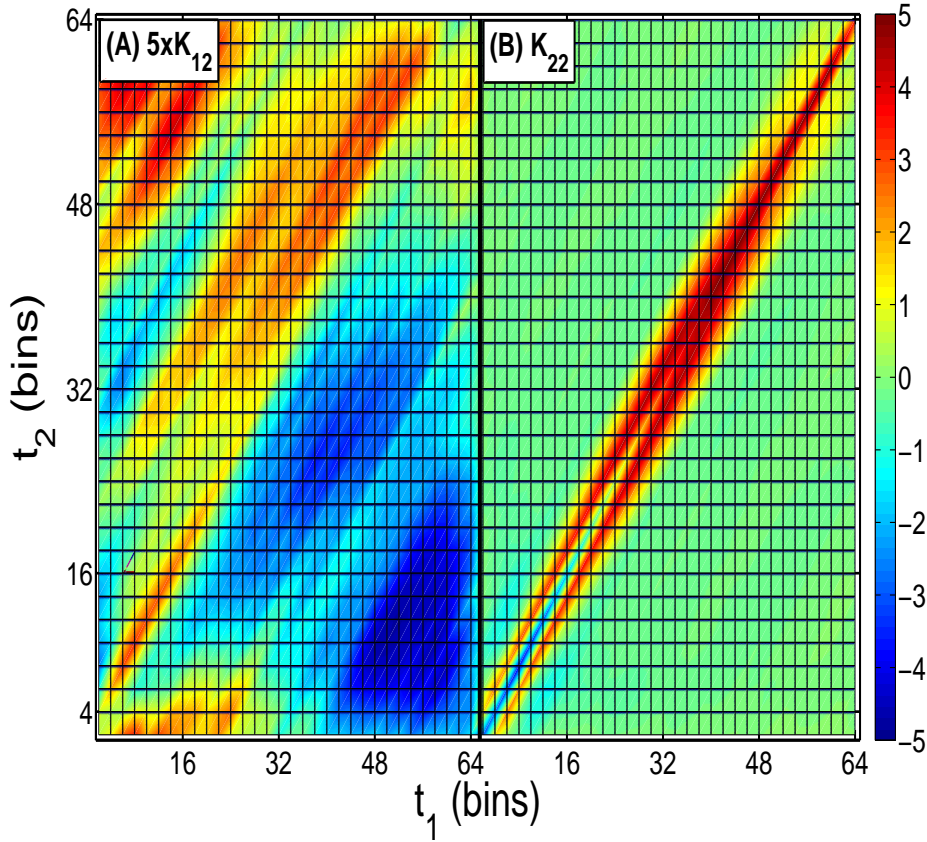,height=12cm,width=14cm}}}
 \caption{ {\sf Second order kernel  $K_{22}(t_1,t_2)$ and upscaled version of $K_{12}(t_1,t_2)$
for $T_w=64$. (A) $5*K_{12}$ , (B)  $K_{22}$. Notice  $K_{22}/K_{12} \sim 5$.
}}
\label{K2_1111}
\end{figure}

The intermediate steps 1 and 2 leading to equation \ref{Dequ1} now increase, since we have to express several 4-point functions in terms of 2-point functions, not all of them being diagonal.
In the particular case of the two H1 neurons though, we may further simplify this system, neglecting $R_{12}$. Its effect\footnote{The effect of $R^{12}$ may  be included perturbatively- see section \ref{Pop}.} is very small indeed, since for rotational stimuli the action of the two neurons is complementary: an exciting stimulus for one neuron is inhibiting for the other - see Figure \ref{R11_R22}.
Although for translational stimuli both neurons  fire nearly synchronously, the dominant peak near $\tau=0$ in $R_{12}$ is absent, since synchrony is not exact.
In the following we  therefore neglect $K_{12}$. As can be seen in Figure \ref{K2_1111}, $K_{12}$ is only $\sim K_{22}/5$.
Since the contributions of $K_{11}$ and $K_{22}$ are already small, $K_{12}$'s
 $1$ \% effect can be safely neglected  for both types of stimuli.

Our equations now decouple and we get two sets identical to equations \ref{Lin}, one for each neuron.

\section{Reconstructing the fly's stimulus and  measuring its quality\hrulefill}
\label{sect:reconstr}
To test the quality of our reconstructions, we
use the data with
$\eta\sim 0.8, \tau = 10$ milliseconds  and  $ \langle r \rangle\sim 80$ spikes sec$^{-1}$.

\begin{figure}[!ht]
 \centerline{\hbox{\psfig{file=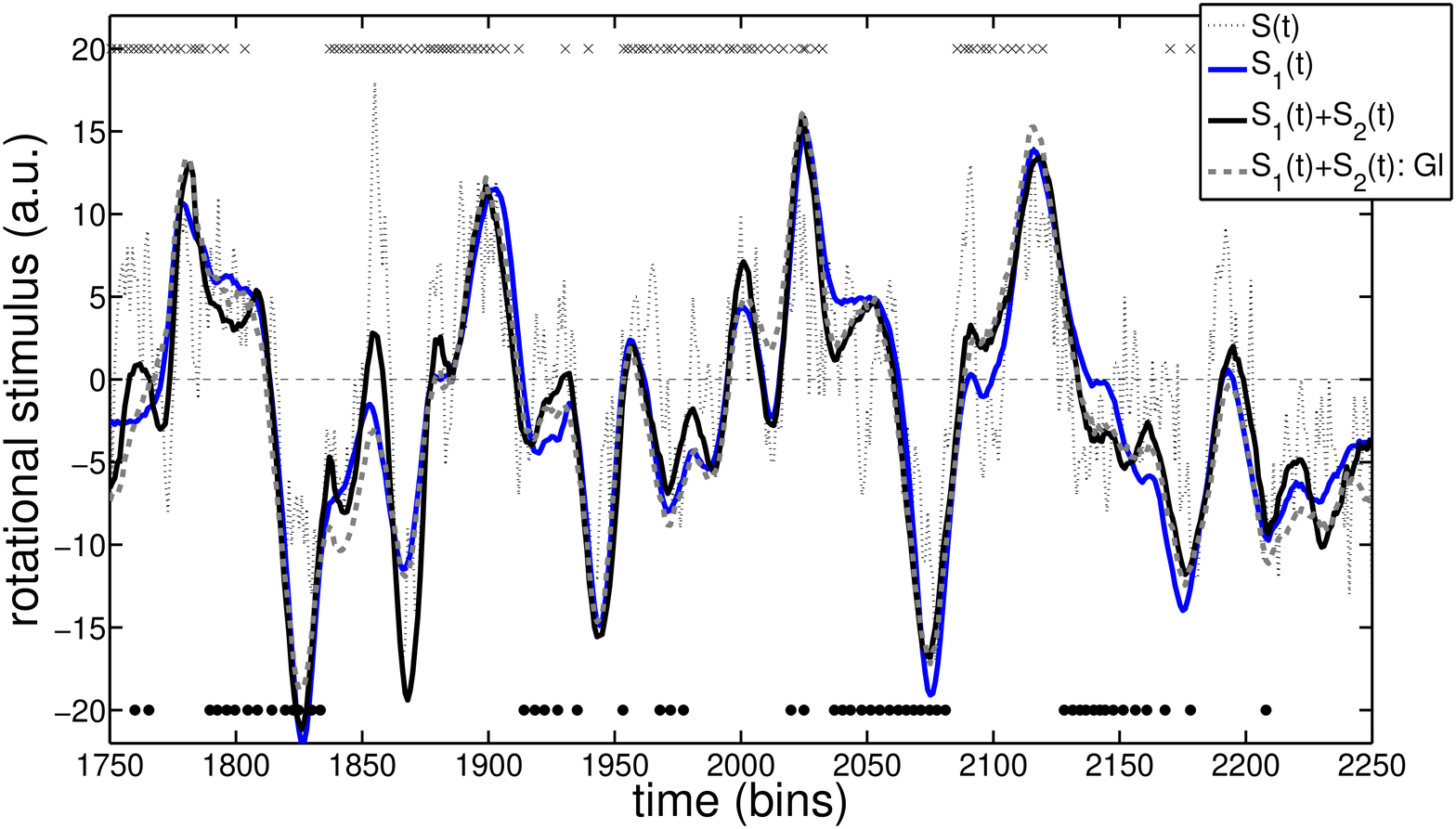,height=8cm,width=16cm}}}
\caption{ {\sf Reconstructing the rotational stimulus with kernels $K_1$, $K_2$ and $K_{11}$, $K_{22}$, using the experimental 4-point function and the Gl-approximation. Black thin dashed line: $S(t)$, input stimulus to be reconstructed, blue line: $S_1(t)$, reconstruction using only $K_1$ and $K_2$, black continuous line: $S_1(t)+S_2(t)$, experimental second order reconstruction, gray dashed line: $S_1(t)+S_2(t): Gl$, Gl-second order reconstruction. The $\times$ and $\bullet$ signs stand for the right and left spikes respectively. Observe a delay-time of about $20$ bins.
}}
\label{Fig:recrot}
\end{figure}

We select a representative sample, one second long,  of the experiment, in order to give a visual display of the reconstruction.
In Figure \ref{Fig:recrot} we show the first order reconstruction of
the original stimulus using $K1$ and $K2$
and the second order reconstruction, where the effect of $K_{11}$ and  $K_{22}$ is added - with and without the Gl-approximation.
We conclude:
\begin{itemize}
\item
Reconstructions using the experimental 4-point functions are very similar to  their Gl-approxi\-mation.
\item
 The reconstruction procedure is unable to
reproduce the fast stimulus variations at the 2 milliseconds time scale.
It is also clear that still higher order terms are not going to improve this deficiency.
But  the second order
kernels always represent an improvement, since the black line in Figure \ref{Fig:recrot} is always a better
approximation to the stimulus than the blue one.
\item
We observe a stimulus-to-spike delay time of $t_{rot}\sim 20$ bins.
\end{itemize}

Although visual appraisement of the reconstruction quality is an indispensable guide to our intuition,
numerical measures are less subjective. We naturally use the $\chi^{(2)} = \langle\int dt [s_e(t)-s(t)]^2 \rangle$ of Equation \ref{chi}, since its minimization was used to determine the
kernels $k_i,K_j$.
The reconstruction improvement due to second order kernel is reflected in
\begin{equation}
       \delta\chi^{(2)} \equiv \frac{ \chi^{(2)}_{1}-\chi^{(2)}_{12}}{\chi^{(2)}_1},
\end{equation}
where $\chi^{(2)}_{1}$ takes only first order terms into account - $\chi^{(2)}_1 = \langle\int dt [s_1(t)-s(t)]^2 \rangle$,
whereas second order terms are included in $\chi^{(2)}_{12} = \langle\int dt [s_1(t)+s_2(t)-s(t)]^2 \rangle$.
$ \delta\chi^{(2)}$ is  positive, but small of  $\sim 8$\%. The chi-squared difference between the experimental and Gl-reconstructions is only of $\sim 0.5$ \%.

\begin{figure}[!ht]
 \centerline{\hbox{\psfig{file=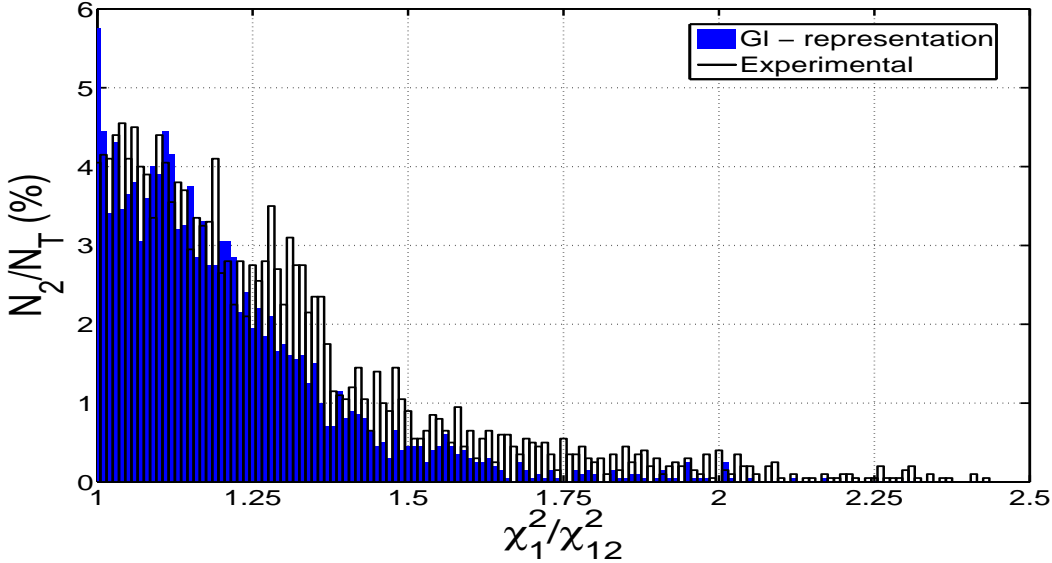,height=8cm,width=16cm}}}
\caption{ {\sf $N_2/N_T$ versus $\chi_{1}^2/\chi_{12}^2$
for experimental and Gl-reconstruction.
We find the instants where $\chi_{1}^2/\chi_{12}^2$ assumes a particular value $\ge 1$, when computed in windows of size $\Delta T=64$ bins. $N_2$ is the number of these windows, whereas $N_T$ is the duration of the experiment in bins divided by the window size.
}}
\label{fig:lchi12}
\end{figure}

Although the $\chi^{(2)}$-improvement is small,  second order terms are a important at specific stimulus-dependent instants. In order to assess the relevance of these, we measure {\em   local chi-squares}, defined as:
\begin{equation}
 \chi_{1}^2(t,\Delta T) \equiv \int_{t-\Delta T}^{t+\Delta T}  dt \langle (s_1(t)-s(t))^2 \rangle
\end{equation}
and
\begin{equation}
\chi_{12}^2(t)\equiv  \int_{t-\Delta T}^{t+\Delta T} dt\langle (s_1(t)+s_2(t)-s(t))^2 ,\rangle
\end{equation}
for $t=T_2$, where $T_2$ are instants when  $\chi_{12}^2(t)$ is at least as important as $\chi_{1}^2(t)$.
If $N_2$ is the number of such windows of size $\Delta T$ and $N_T$ the duration  of the experiment in bins divided by the window-size in bins, we
plot in  Figure \ref{fig:lchi12}  the fraction of the stimulus-dependent instants vs.
$\chi_{1}^2/\chi_{12}^2$.
Although this fraction vanishes as we require the importance of  second order terms to increase, they still make a sizable contribution.
Unfortunately just looking at the mean stimulus around $T_2$ does not provide any insight
 and a more detailed analysis will be needed to reveal features, which might be relevant at these particular instants.

Here we only follow \cite{Spikes} and separate systematic from random errors,
decomposing the estimate $\tilde s_e(\omega)$  into a frequency-dependent gain $g(\omega)$ and an
effective noise $n_{eff}(\omega)$ referred to the input:
\begin{equation}
	\tilde s_e(\omega) = g(\omega)[\tilde s(\omega)+n_{eff}(\omega)].
\end{equation}
Around $T_2$, we observe an overall improvement of 20\% in $g(\omega)$. A further indication,
that second order contributions, although drowned in averages over the whole
experiment, may nevertheless have crucial importance in  improving the code at specific moments.

\begin{figure}[!ht]
 \centerline{\hbox{\psfig{file=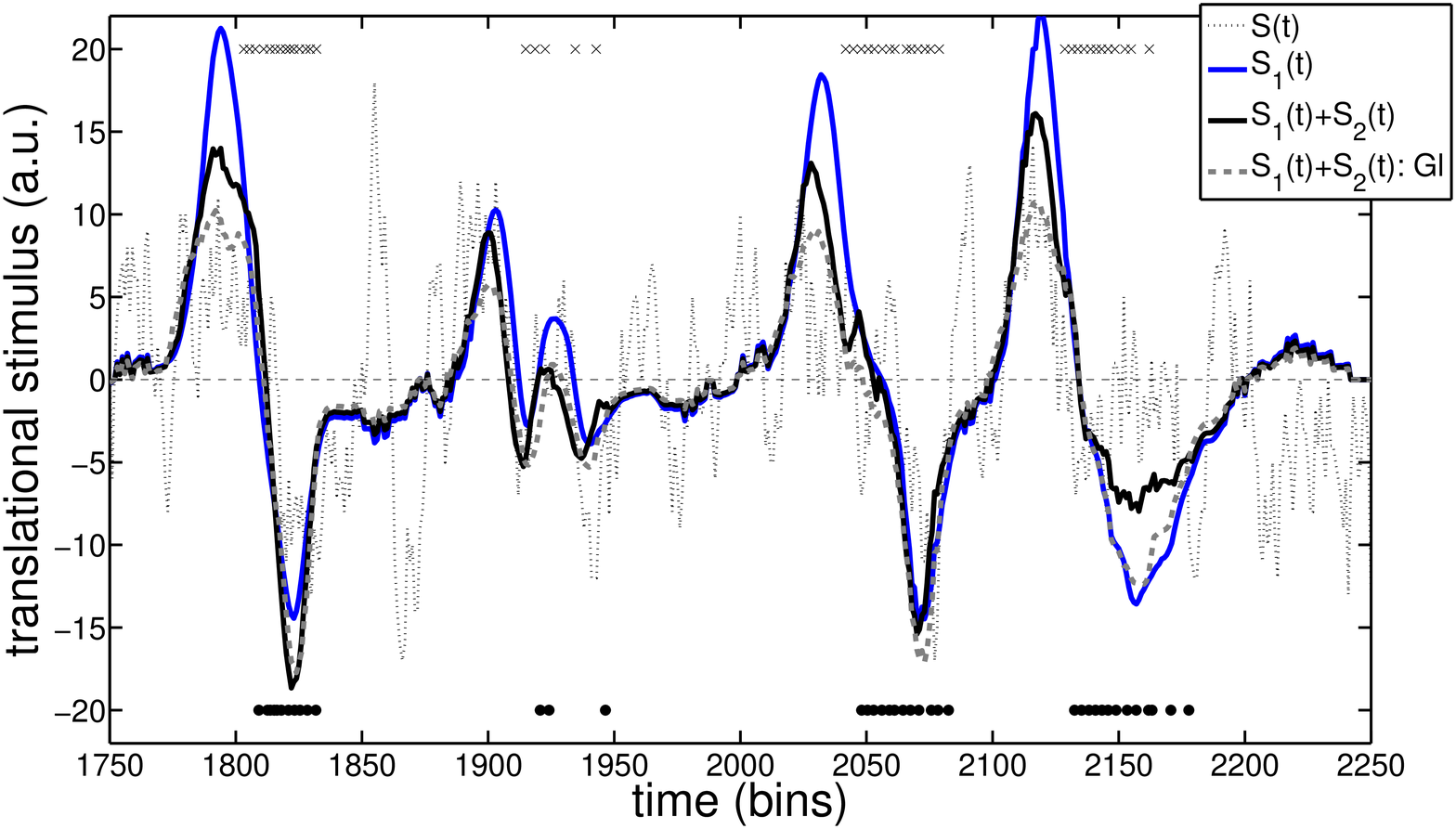,height=8cm,width=16cm}}}
\caption{ {\sf Reconstructing the translational stimulus with kernels $K_1$, $K_2$ and $K_{11}$, $K_{22}$, using the experimental 4-point function and the Gl-approximation. Black thin dashed line: $S(t)$, input stimulus to be reconstructed, blue line: $S_1(t)$, reconstruction using only $K_1$ and $K_2$, black continuous line: $S_1(t)+S_2(t)$, experimental second order reconstruction, gray dashed line: $S_1(t)+S_2(t): Gl$, Gl-second order reconstruction. The $\times$ and $\bullet$ signs stand for the right and left spikes respectively. Observe a delay-time of about $25$ bins.
}}
\label{Fig:rectrans}
\end{figure}

Finally we  discuss the reconstruction of translational stimuli.
Although in real life there is a continuous intermingling of {\em rotational} and {\em translational} motion, for a start we have considered this artificial
separation of  stimuli. Thus  we have computed all averages $\langle \cdot \rangle$ also for the translational setup.
The kernels $K_a,K_{ab}$ are similar to the rotational ones,
but there is a sign change. Whereas for rotational stimuli $K_1\sim -K_2, K_{11}\sim -K_{22}$, for the translational case we have
\begin{equation}
\begin{array}{ccc}
{K_1}^{(trans)} & \sim {K_2}^{(trans)}& \sim  {K_1}^{(rot) } ,\\
{K_{11}}^{(trans)}&\sim  {K_{22}}^{(trans)}& \sim  {K_{11}}^{(rot) }  ,
\end{array}
\end{equation}
The reconstructions shown in Figure \ref{Fig:rectrans} are worse than the rotational ones. For positive stimuli, corresponding to unrealistic backward motion of the fly, both neurons fire
vigorously, whereas in the opposite case none does. Interestingly, the delay-time is now $t_{trans}\sim 25$ bins, about $5$ bins larger than $t_{rot}$: inspite of their mutual inhibition, the neurons manage to fire, albeit a little bit retarded.
The Gl-representation works equally well for this case.
It would be interesting to subject the fly to a more realistic mixture of rotational and
translational motion without separating the two and then compute correlation functions etc.
We intend to come back to this issue in the future.

\section{Gl-approximation in population coding: taming the matrix explosion\hrulefill}
\label{Pop}

Although the spike generation process of the H1 neurons is not Gaussian, the
parametrization  \ref{Fit1} is unexpectedly good. Actually we don't know how
to judge from the spike interval distribution, whether this surprise will happen or not.
In fact, the  interval distribution of the spike times looks more nearly Poisson, instead of Gaussian. We remark, that independent
increment probability distributions, whether they are Poisson or not, never do justice to correlated spike trains.
On the other hand, if the 2-point function $R(t)$ is to be a suitable building block to
represent the 4-point function, then the parametrization, equation \ref{Fit1}, is uniquely
selected to be the most general one respecting the symmetry of $R^{(4)}(1,2,3,4)$.

Since first order computations treat each neuron independently and do not take their mutual correlations into account, in the future one certainly would want to perform second  order reconstructions
to study the fly's visual system for more than two neurons. Our Gl-approximation makes these computations much more feasible. It should also work for correlation functions involving  neurons not belonging to the fly's lobula plate.

In order to apply our Gl-approximation, we imposed  the requirement $R_{11} = R_{22}$ and
we neglected $R_{12}$.
This limitation may be relaxed in the following way\footnote{Here we only provided an outline, leaving a detailed analysis for a  future publication.}.
One could set $R_{12}=0$ and use a different set of functions for each neuron, diagonalizing thus all
2-point functions ${\cal R}_{aa}$ and compute the coefficients ${\cal D}_{ab}$. Then reexpand all variables in
terms of one set of functions only and
 apply the procedure, which led to equation \ref{EquD1} for $R_{12}\neq 0$.
If this does not lead to a closed set of equations, small effects may always
be taken into account by a perturbative scheme to arbitrary order.
In fact, suppose we have solved equation \ref{Lin} for some representation of ${\cal R}^{\mu\nu\alpha\beta}_{abcd}$ - e.g. as we did in section \ref{1G}.
Incorporating  $R_{12}\neq 0$ and/or $R_{22}^{\mu\nu}\neq\delta_{\mu\nu}$ will change the ${\cal R}$-matrix into:
\begin{equation}
{\cal R'\,} = {\cal R}+
\delta{\cal R},
\end{equation}
with  $\delta{\cal R}$ supposedly {\em small}.
The new equations to be solved are:
\begin{equation}
{\cal S}_{ab}^{\mu\nu} =
\sum_{cd,\alpha\beta} {\cal R'\,}_{abcd}^{\mu\nu\alpha\beta}\, {\cal D'\,}_{cd}^{\alpha\beta},
\label{Lin2}
\end{equation}
where ${\cal D'}={\cal D}+\delta{\cal D}$ and ${\cal D}$ satisfies
the  unprimed equations \ref{Lin}.
Expanding both sides of equation \ref{Lin2}  to first order in the corrections, we get the equations
\begin{equation}
-\sum_{cd,\alpha\beta}\delta {\cal R}_{abcd}^{\mu\nu\alpha\beta}\,{\cal D}_{cd}^{\alpha\beta} =
\sum_{cd,\alpha\beta} {\cal R}_{abcd}^{\mu\nu\alpha\beta}\,\delta {\cal D}_{cd}^{\alpha\beta}
\label{Lin3},
\end{equation}
to be solved for the unknowns $ \delta {\cal D}$.
$(-\delta {\cal R}\cdot{\cal D})$ replaces the left-hand-side of equation \ref{Lin} and couples the neurons.
The right-hand-sides of the above equation and  equation \ref{Lin} have the  same form and can therefore be solved in the same manner.

The Gl-approximation  could also be useful for other systems and this would be a considerable step forward in implementing
 coding involving  a large population of neurons.
One of the problems in second order reconstructions involving many neurons is
the size-explosion of the 4-point function matrices alluded to at equation \ref{4ptmat}.
If, e.g. we record from four neurons using $128$ bin-sized windows, the length of the matrices to be inverted would be
$\sim 128^8 \times 2^8 \sim 10^{19}$. With our approximation the size of the linear system  to
be solved
grows only linearly with the number of neurons.

In order to use our approximation, one would have to check
the win\-dow\-size independence of the parameters $A_{ab\ldots}$ and $B_{ab\ldots}$ for some subset of
the complete matrix-indices, to convince oneself of the adequacy of the approximation. Since in our case
the matrices were still manageable, we could compute the experimental 4-point functions to verify this point, but this will in general not be possible.

\section{Materials and Methods \hrulefill}
\label{MM}

Flies, immobilized with wax, viewed two Tektronix 608 Monitors M1, M2, one
for each eye, from a distance of $12cm$, as depicted in
Figure \ref{setup}. The monitors were
horizontally centered, such that the mean spiking rates of the two neurons,
averaged over several minutes, were equal.
They were positioned, such that a straight line connecting the most sensitive spot of the compound eye to the monitor was perpendicular to the monitor's screen.
The light intensity corresponds roughly to that seen by a fly at dusk  \cite{rob:1997}.
The stimulus was a rigidly moving vertical bar pattern with horizontal velocity $v(t)$.
We discretise time in bins of 2 milliseconds,
which is roughly the refractory period of the H1 neurons. The fly therefore
saw a new frame on the monitor every $\delta t = 2$ milliseconds,
whose change in position $ \delta x$ was given by $\delta x(t) =
v(t) \delta t$.

The velocity $v(t)$ was generated by an Ornstein-Uhlenbeck process with correlation
times $\tau_c = 0, 5$  and $10$ ms \footnote{Although we show results only for $\tau_c=10$ ms our conclusions are also valid for $\tau_c = 0, 5$  ms.},
i.e. the stimulus was taken from a Gaussian
distribution with correlation function $C(t) = e^{-t/\tau_c}$. Experimental runs for each $\tau_c$
lasted 45 minutes, consisting of 20 seconds long segments. In each segment, in the first 10 seconds the same stimulus was shown, whereas in  the next 10 seconds the fly saw different stimuli.

\section{Summary \hrulefill}

The ability to reconstruct  stimuli from the output of sensory neurons is a basic step in
understanding how sensory systems operate.
If intra- and inter-neuron correlations between the spikes emitted by  neurons are to be taken into account,
going beyond first order reconstructions is mandatory. In this case one has to face the size-explosion of higher order
spike-spike correlation functions, the simplest being the 4-point correlation function necessary for a second order
reconstruction. Our Gl-representation of the 4-point function in terms of 2-point functions tames this problem.
If this representation holds more generally, the  coding in large  populations would become more feasible.

For  our case of the two H1 neurons of the fly,  correlations between them may be neglected,  since they are only of $\sim 1$ \%.
We perform reconstructions using both the experimental and the Gl-approximation for the 4-point
functions involved. Both are very similar,  their chi-squared differing by $0.5$ \%.
To  implement the  Gl-program for the two neurons, we found it convenient to expand our
variables in terms of eigenfunctions of 2-point matrices.
We propose  a perturbative scheme in order to take the neglected correlations into account.

We find that second order terms always improve the reconstruction, although measured by a chi-squared averaged over
the whole experiment this improvement is  only at the  8\% level. Yet these terms can represent a $100$ \% improvement  at special instants
as measured by an instant dependent chi-squared.\\

\noindent
{\bf\Large Acknowledgments \hrulefill}\\\\
We thank I. Zuccoloto
for her help with the experiments.
The laboratory was partially funded by FAPESP grant 0203565-4.
NMF and BDLP were supported by FAPESP fellowships.
We thank Altera Corporation for their University program and Scilab for its excellent software.

\bibliographystyle{apacitex}

\end{document}